 \pdfoutput=1
\documentclass{vgtc}
\ifx\ifpdf\undefined\else\let\ifpdf\relax\fi

\usepackage{mathptmx}
\usepackage{graphicx}
\usepackage{times}
\usepackage{todonotes}

\onlineid{0}

\vgtccategory{Research}

\vgtcinsertpkg

\title{3D Character Customization for Non-Professional Users in Handheld Augmented Reality}

\author{Iris Seidinger\\
        \scriptsize Salzburg University of Applied Sciences %
\and Jens Grubert\thanks{e-mail: jg@jensgrubert.de}\\ %
     \scriptsize University of Passau %
}

\abstract{In gaming, customizing individual characters, can create personal bonds between players and their characters. Hence, character customization is a standard component in many games. While mobile Augmented Reality (AR) games become popular, to date, no 3D character editor for AR games exists. We investigate the feasibility of 3D character customization for smartphone-based AR in an iterative design process. 

Specifically, we present findings from creating AR prototypes in a handheld AR setting. In a first user study, we found that a tangible AR prototype resulted in higher hedonistic measures than a camera-based approach. In a follow up study, we compared the tangible AR prototype with a non-AR touchscreen version for selection, scaling, translation and rotation tasks in a 3D character customization setting. The tangible AR version resulted in significantly better results for stimulation and novelty measures than the non-AR version. At the same time, it maintained a proficient level in pragmatic measures such as accuracy and efficiency.

} 

\CCScatlist{ 
  \CCScat{K.6.1}{Management of Computing and Information Systems}%
{Project and People Management}{Life Cycle};
  \CCScat{K.7.m}{The Computing Profession}{Miscellaneous}{Ethics}
}

\begin{document}


\firstsection{Introduction}

\maketitle

Despite of the growing popularity of Augmented Reality (AR) games \cite{thomas2012survey}, there has been no character editor in the field of AR, yet. Furthermore, mobile games generally often lack customization and offer only one or a small set of predefined characters to choose from. One reason for that is that such an editor is often unnecessary for games that don't afford much time and commitment of the player. However, with mobile AR games becoming more common, the variety will increase and next to casual games (e.g., \cite{grubert2012playing,grubert2013playing}), 
which are only played occasionally, bigger and more immersive games, focusing on stories and characters, are likely to evolve \cite{thomas2012survey}. This can increase the demand for a character editor, as, by customizing an individual character, people tend to identify themselves more with it and strong personal bonds can be developed \cite{turkay2014effects}. 

The objective of this work is to investigate the potentials of handheld AR environments for character customization. To this end, we iteratively designed, implemented and validated a series of mobile AR prototypes. In contrast to professional character modelling, we focus on non-professional users who want to quickly customize a game character before playing a game. Hence, we focus on constrained customization of 3D characters using 3D object selection and manipulation tasks.

\section{Related Work}

Our contributions are based upon the rich body of work in 3D modelling and 3D user interfaces. 3D modelling applications can implement various functions, input methods and interaction designs. However, most of them share some basic functions that can be grouped together into object manipulation, viewpoint manipulation and application control \cite{hand1997survey}. 

There is a large body of relevant work in 3D User Interfaces for Virtual and Augmented Reality scenarios (for an overview see \cite{bowman20083d}). Here, we focus on object selection and manipulation in mobile AR. Interaction techniques typically rely on camera-based input  (e.g., \cite{henrysson2007experiments,harviainen2009camera}, also in combination with touch input \cite{bai2012freeze,marzo2014combining,tiefenbacher2014poster}), mid-air hand or finger input (e.g., \cite{hurst2013gesture,buchmann2004fingartips,bai2012freeze}), tangible objects (e.g., \cite{kato2000virtual,billinghurst2009advanced,bai2013free, sodhi2013bethere}) or a combination thereof.

\emph{Camera control} is challenging in handheld AR as the camera can be manipulated with 6 degrees of freedom but it is typically hard for users to accurately and precisely align the device with a specific pose. Approaches for supporting the user in this tasks encompass, e.g., attention funnels \cite{biocca2006attention}, iron sight visualizations \cite{hartl2013mobile, hartl2016efficient} or freezing the camera view \cite{lee2009freeze, grubert2012exploring}. For \emph{object selection}, applications often use a ray casting metaphor. Examples include 3DTouch and HOMER-S \cite{mossel20133dtouch}, TransVision \cite{rekimoto1996transvision}, the Personal Interaction Panel \cite{szalavari1997personal} and Napkin Sketch \cite{xin2008napkin}. DrillSample \cite{mossel2013drillsample} also uses the pointing metaphor by interpreting taps on a smartphone screen, but extends the selection process by adding the so-called DrillSample view to display all objects, that could be selected by this tap, including all hits instead of only the first one. Gesture-based selection of objects has also been investigated, e.g., employing finger tracking and pinching two fingers together above an object to select it (e.g., \cite{hurst2013gesture,buchmann2004fingartips}), bringing the original HeadCrusher idea \cite{pierce1997image} to mobile AR. Lee et al. propose to use marker occlussion to select object \cite{lee2004occlusion}, similar to Vuforia's virtual button technique\footnote{https://developer.vuforia.com/}. Also, tangible user interfaces (TUIs) were explored for object selection (and manipulation), such as VOMAR \cite{kato2000virtual} or project Magic Cup \cite{billinghurst2009advanced}. For \emph{translation and scaling tasks}, it is common to allow free positioning in 3D by attaching the object to a tool or the tracked hand. Examples include the Personal Interaction Panel \cite{szalavari1997personal}, FingARtips \cite{buchmann2004fingartips}, HandyAR \cite{lee2007handy}, Magic Cup \cite{billinghurst2009advanced},  VOMAR \cite{kato2000virtual} or, recently, work by Kim et al. \cite{kim2016touch}. In HOMER-S \cite{mossel20133dtouch}, the object doesn't get locked to a tool or hand, but instead locks its relative position directly to the device's camera. The user can therefore move the device around and to the desired location considering the relative positioning of the object to the camera. The combination of multi-touch and device movements was also investigated \cite{marzo2014combining,tiefenbacher2014poster}. For \emph{rotation tasks}, approaches commonly rely on constraint rotation around individual axes (e.g., \cite{au2012multitouch, billinghurst2009advanced, mossel20133dtouch, hurst2013gesture} and some on free rotation by attaching the object to a hand or prop  (e.g., \cite{kato2000virtual, buchmann2004fingartips}.


\section{Initial Interactive Prototypes}

We initially investigated both a camera-based approach and a tangible AR system for manipulating individual body parts of a 3D character (head, arms, torso, legs). An important requirement when developing the prototypes was to strive for an entertaining experience while maintaining a satisfying precision and efficiency.  Since a character editor is part of a game, people are using it with the intention of having fun with the game. Therefore, it is a good idea to make the process of creating the character as fun and entertaining as possible. It should also be noted that most games only offer character customization instead of actual character creation. This is because games typically rely on the character to follow certain restrictions. Hence, we restricted the prototypes on standard spatial manipulation tasks for certain body parts. In the first iteration, we only focused on selection and scale tasks, while in the final AR prototype we also included translation and rotation.



\subsection{Camera-based approach}
In the initial camera-based approach the 3D pose of a handheld device is used both for viewport manipulation and object manipulation  (similar to \cite{harviainen2009camera}). Object selection is done through raycasting (the user touches the projection of a 3D object on screen). For scaling, the device is moved closer or further towards the object and the selected body part is enlarged as the device is moving away or shrunk of the device moves closer. When scaling, all body parts automatically adjust their positions in a way that their point of intersection with the body remain unchanged. A clutch mechanism (using an on-screen button) is used for activating the action.




\subsection{Tangible UI}

\begin{figure}[!htbp]
\centering
\vspace*{0.3in}
    \includegraphics[width=0.8\columnwidth]{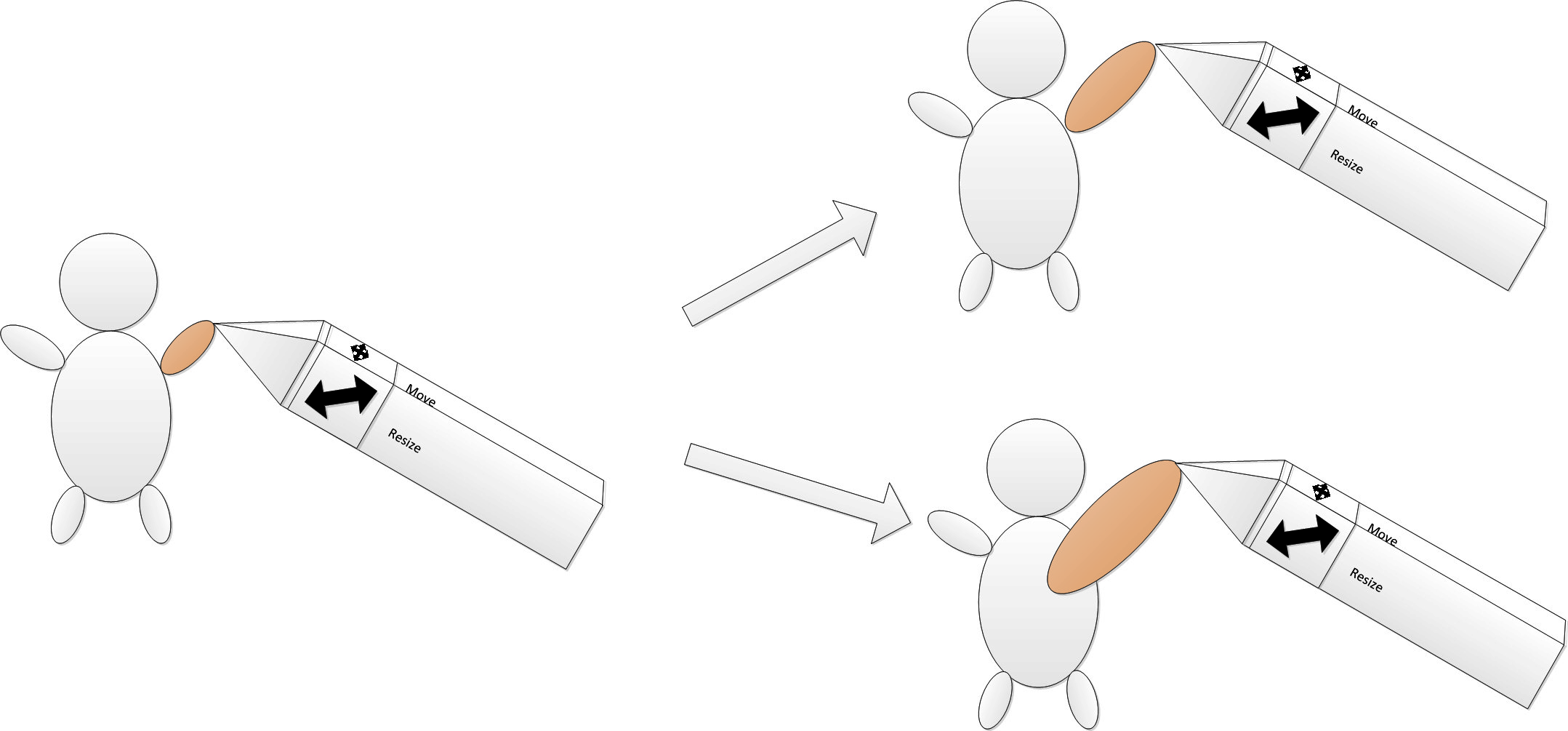}
\vspace*{0.2in}
    \caption{MarkerPen-based approach: Interaction steps for scaling:
    1) Arm is preselected and scale marker is tracked 2) the MarkerPen is moved and the arm is resized to touch the pen's tip. Top: with directional scaling, the arm keeps its relative position to the torso. Bottom: without directional scaling}
    \label{fig:interSCA}
\end{figure}

We used a pen metaphor for spatial manipulation of character parts. The device could be used for interactions using a pointing metaphor, where the pen tip is the pointer. Rotating the pen to a side with a different marker could trigger a new manipulation mode, which additionally has to be activated by pressing a lock button on the touchscreen. For the scaling interaction, as visualized in Figure \ref{fig:interSCA}, the associated scale marker must be tracked by the device's camera. By this, the scale mode is activated and the user can start the manipulation phase by pressing the lock button. The selected object will now always scale to a size, where its hull touches the MarkerPen's tip. 
When the user moves the MarkerPen away from the object, the object increases in size, following the pen tip and when tip is moved towards or into the object, it scales down to again match the size pointed out by the tip. By pressing the lock button again, the object is released from the manipulation mode and the scale is applied. 


\vspace{-0.5cm}

\subsection{Initial Evaluation}

The first evaluation compared two early-stage prototypes of the camera-based approach and the MarkerPen version for scaling an object. A repeated-measures design was used. The subjects could test both approaches, filled out the user experience questionnaire (UEQ) \cite{laugwitz2008construction} and gave qualitative feedback as part of a semi-structured interview. The target of this iteration was to get a  impression of how the users are interacting with and responding to the prototypes, if the prototypes are suitable for the task and which improvements might be necessary. Finally, one protoype should be selected for further development. 

Six students with background in computer science participated in the study (1 female, 5 male, mean age  25.6 years sd=3.4).

The test application for the initial evaluation cycle was built Unity3D and consisted of a simple menu, from which either prototype could be launched, and the prototypes themselves. This application was installed on a HTC One M7 running Android 5.0.2. Next to the actual prototype application, a screen and microphone recorder was running to create audio and video material for later analysis and reference. Each prototype showed the same model which consisted out of a cube as body and a sphere as head with some basic textures. No selection was possible, the head was preselected and the only possible interaction was the resizing of the head. The evaluations took place in a laboratory environment. 

At the beginning of the evaluation, every participant was given a short informal introduction on the procedure of the experiment and they were asked if they agree to the voice recording, which all of them did. Following the concept of a think-aloud protocol 
, participants were asked to continuously express their thoughts to gain further insights into users' expectations and understanding of the user interface. After this introduction, voice recording was started and the test person was asked to try out the first prototype. No time limit was given to the participants so they could experiment with the prototype as long as they wanted to. Most of them tested each prototype for about two to five minutes while expressing opinions, comments and ideas for improvement. The participants were then asked to fill the UEQ questionnaire. After they finished testing the first prototype, the test setup was changed according to the prototype (e.g. positioning the marker correctly, preparing the mounting, ...) and the participant was asked to start testing the other prototype. During the evaluation, the probands were allowed to ask questions or ask for help if needed. The starting order of the applications was randomized.

\subsection{Results and Discussion}

\begin{figure}[!htbp]
    \centering
\includegraphics[width=1\columnwidth]{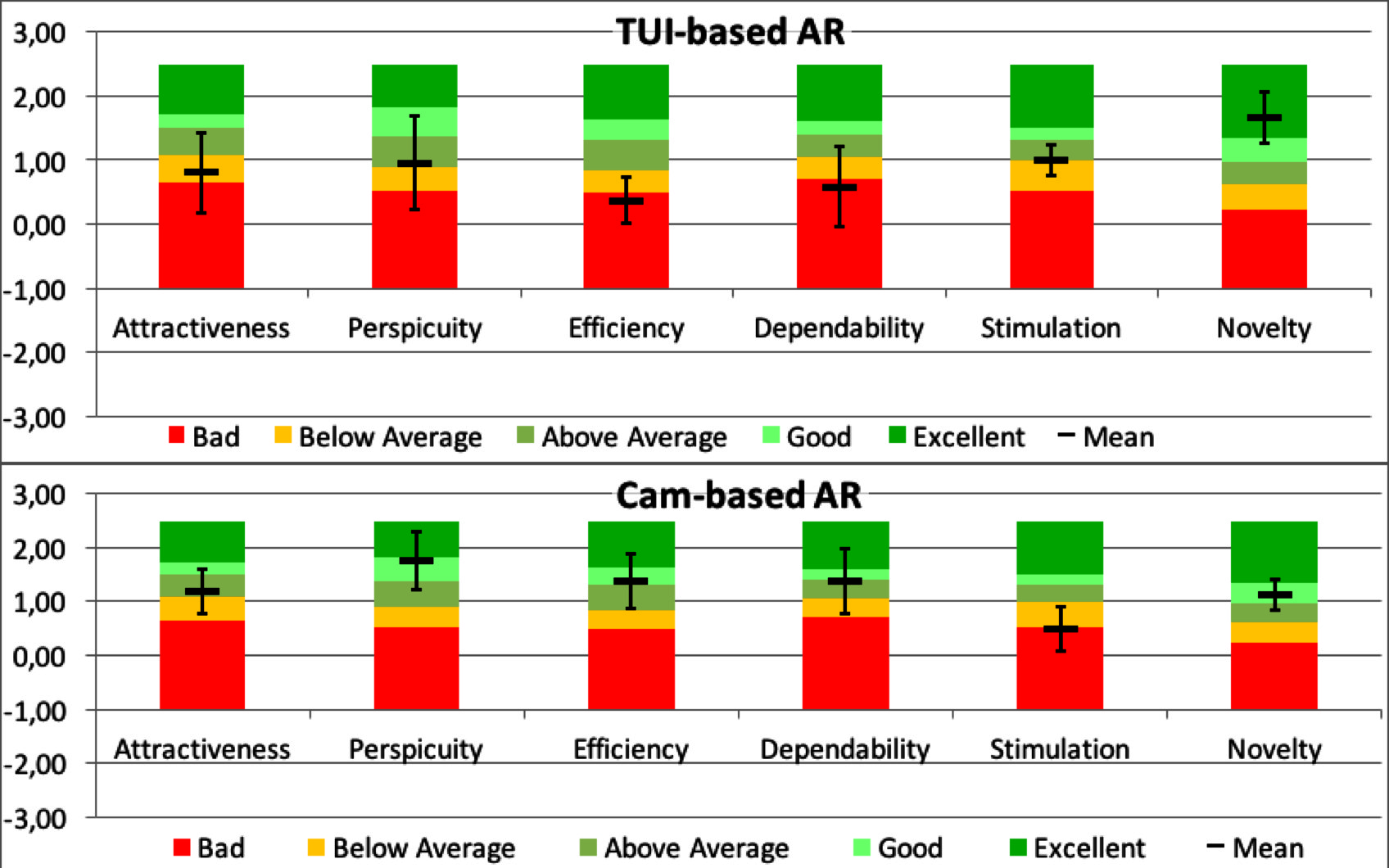}
    \caption{Results of the UEQ for the TUI-based prototype (top row) and camera-based prototype (bottom row) in the inital evaluation.}
    \label{fig:ueq1}
\end{figure}

Figure \ref{fig:ueq1} shows the results from the UEQ questionnaire. The values represent the means of the scales, ranging from -3 to +3 where -3 means an extremely bad performance and +3 an extraordinary high performance. The camera-based prototype scored high in perspicuity, efficiency and dependability while scoring lower on attractiveness, stimulation and novelty. The figure also shows how well the camera-based prototype performed in relation to other products and innovations from the UEQ benchmark data set. The benchmark includes data from 4818 persons from 163 studies about various products and "allows conclusions about the relative quality of the evaluated product compared to other products" \cite{ueq}. Findings from qualitative analysis show that the camera-based prototype was well received for its practical and functional aspects. Participants described it as easy to understand, that it worked well and that the tracking was good. 

Compared to other products from the UEQ benchmark, the MarkerPen prototype scored well in stimulation and novelty, but only scored average and below average for the other measures. The approach scored lowest for efficiency and dependability. The mean of the dependability scale for the MarkerPen prototype has a high deviation. As the tracking of the frame marker  was prone to errors and easily affected by changing lighting conditions, it is possible that this scale could improve substantially when the frame marker tracking of the prototype is further improved. Some participants had problems with learning the interaction with the MarkerPen. They stated to not feel in control and having problems with holding the pen correctly. This seemed to improve when they get accustomed to the interaction. Then, those participants found the prototype useful and even intuitive and thought the interaction "feels good". 

When comparing the UEQ findings of the camera-based and the MarkerPen prototype, one prominent difference shows in the performance in hedonistic and pragmatic measures. While the camera-based approach scores high on pragmatic quality but lacks hedonic quality, the MarkerPen-based prototype scores high on hedonic, but low on pragmatic quality. No significance tests were made due to the small number of participants. Furthermore, there was no necessity for it, as the evaluation was primarily conducted to gain information for future prototype development decisions.

This research aims to find a suitable interaction design for a 3D AR character editing tool for games. For the goal of providing simple interactions that allow fast as well as more detailed modeling, both prototypes provided a satisfying experience, however the camera-based approach was easier to learn and it outranked the MarkerPen prototype in terms of efficiency. It could not be inferred at this stage if fast prototyping as well as detailed modeling  are easily supported by the prototypes. An important requirement, is to strive for an entertaining experience, as long as a satisfying degree of precision and efficiency can still be provided.
Since the hedonic quality of the MarkerPen-based approach is substantially higher than that of the camera-based approach, and due to the assumption that the pragmatic measures of the MarkerPen prototype will improve with further development and enhancements of the frame marker tracking, the MarkerPen approach was developed further.


\section{Refined prototypes and evaluation}

The final evaluation compared a refined version of the  MarkerPen prototype  to a Non-AR prototype. The aim of this evaluation was to determine how well the developed AR prototype performs relative to a traditional approach to mobile character editing. The final evaluation again used a repeated-measures design, with counterbalanced starting order of the prototypes.  

\subsection{Refined tangible UI}
\begin{figure}[!htbp]   
    \centering
    \includegraphics[width=0.95\columnwidth]{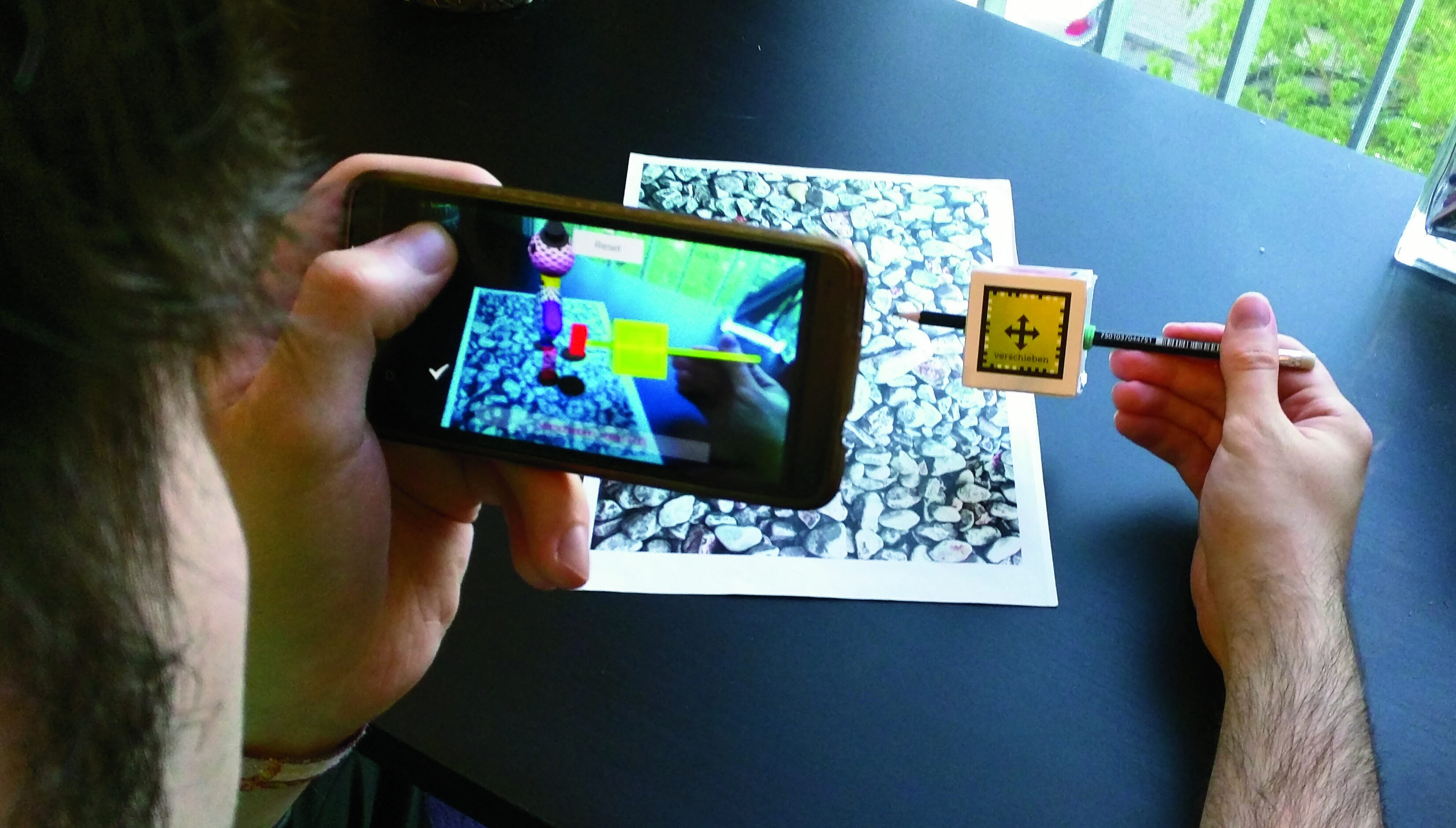}\\
    \includegraphics[width=0.95\columnwidth]{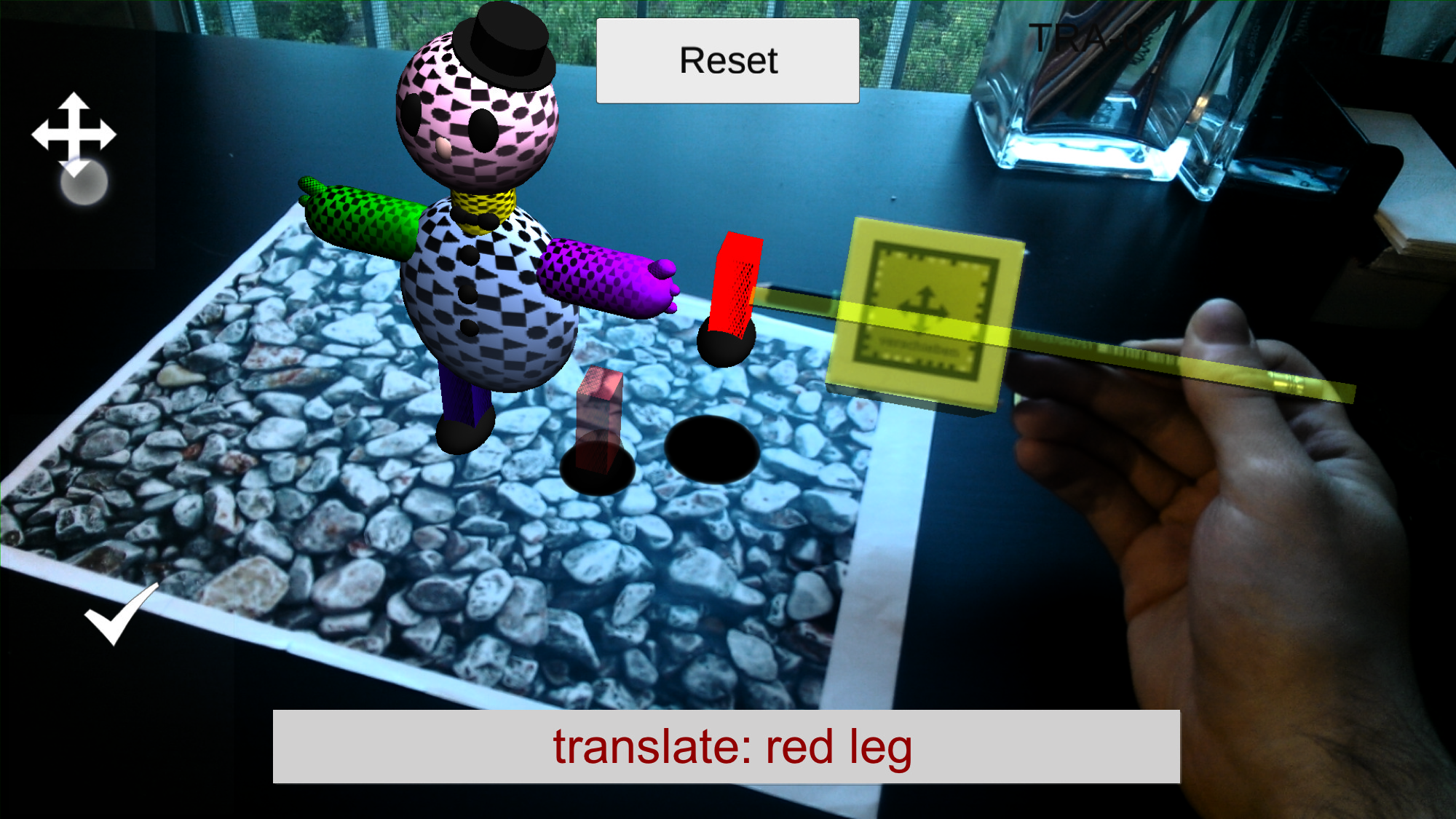}
    \caption[Final prototype of the MarkerPen in use ]{Final prototype of the MarkerPen in use (top row) and screenshot of application (bottom row)}
    \label{fig:markerpenFinal2}
\end{figure}

The final prototype was built similarly to the previous one. This time however, foam board was used for the marker cube instead of cardboard, to provide a better stability. Also, a hexagonal pencil was used instead of a round one and hexagonal holes were cut into the foam board cube to stick the pencil through. This prevented the marker cube from spinning around the pencil. Figure \ref{fig:markerpenFinal2} shows the AR prototype in use. 

Four different frame markers were created for the prototype, one for each manipulation mode. Each of them features its own color, a matching icon in its center and a descriptive verb beneath it to make it easier to understand and distinguish between the modes. Each marker was attached to one side of the foam board cube. The on-screen mode trigger button for starting and stopping manipulations, which was introduced with the first prototypes worked well but the naming ("Lock"/"Release") and appearance confused the users in the first place. Therefore, the button was improved in the final prototype and is now only visible if the MarkerPen is tracked. Furthermore, it displays the same icon as can be seen on the tracked operation marker. This emphasises the relationship between the tracking of a mode marker and the button on the touchscreen. Updating the buttons icon depending on the tracked operation marker also helps the user to understand what the mode trigger does. As depth perception can be a challenge in AR applications, a shadow for the MarkerPen's tip was added to give the user a better understanding of the pen's location in depth. 


\subsection{Non-AR comparison application}
\begin{figure}[!htbp]   
    \centering
    \includegraphics[width=0.9\columnwidth]{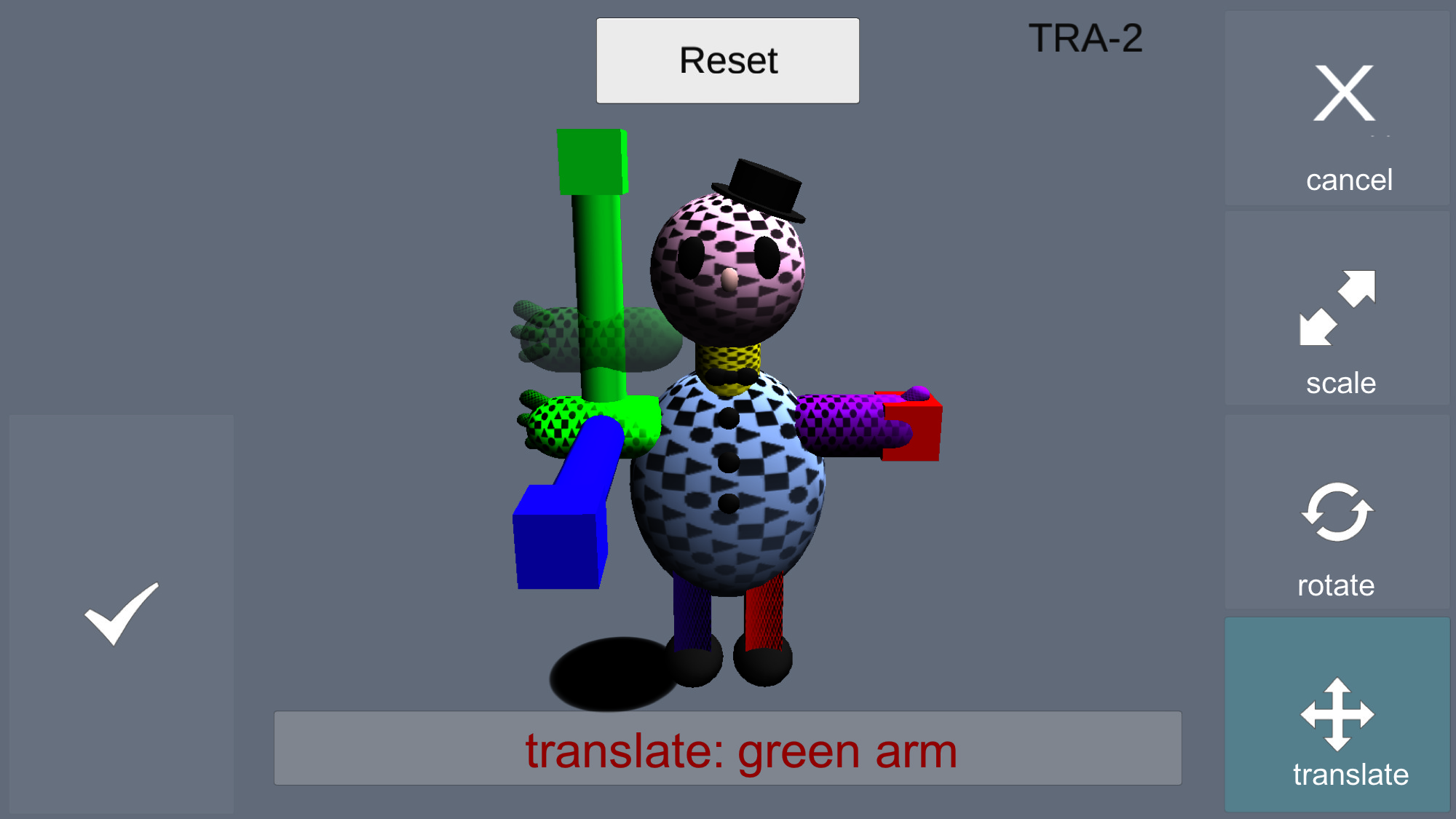}
    \caption{Screenshot of the Non-AR prototype. The current task is displayed in red, the selected action is highlighted in dark green, the target pose is indicated by a semi-transparent copy of the body part to be manipulated.}
    \label{fig:narScreen}
\end{figure}

We compared a number of mobile 3D editing applications (Creationist 3D modeling, Qubism 3D modeling FormIt 360, Sketcher 3D, Spacedraw). However, none of them allowed the implementation of target objects that could be used to visualize the target manipulation of an object to solve spatial manipulation tasks. To this end, an existing in-house character editing software was adapted to meet the requirements of this research. This prototypical software aimed to implement transform handles similar to the ones used in 3D Creationist\footnote{http://www.3dcreationist.com/} 
that could be applied to any 3D object. It includes the basic transformations rotation, translation and scale and just like in 3D Creationist, the handle bars can be dragged to apply those manipulations along that axis, see Figure \ref{fig:narScreen}.

\subsection{Participants}
For the final evaluation, ten probands of different professions and experiences with smart phones and gaming participated in the study (4 female, 6 male, age: 22-29 years).  Eight test persons stated that they are experienced or very experienced in the use of smartphones and nine play games on their mobile devices at least occasionally. This matches the target group of this research, which consists of intermediate and advanced smartphone users interested in mobile gaming.  Five probands had experience in professional 3D modeling and seven already had experiences with AR. Two persons indicated to be very experienced with AR applications, and stated that they used AR apps frequently. Nine of ten participants used a character editor at least once before, eight of them used character editors more than once. 

\subsection{Hypotheses}
Our first hypothesis was \textit{H1: Touch-based manipulation will be faster than the MarkerPen approach}. This hypothesis originates from the fact that it costs some time to rotate the MarkerPen and (re-)track the markers. Additionally, handling and coordinating the MarkerPen and the smart phone probably costs more time than only interacting with the touch screen, where only small movements are required. Tracking problems can affect the accuracy and efficiency, therefore the second hypothesis was 
Hence, the second hypothesis was \textit{H2: MarkerPen-based manipulation will be more exciting and fun to use than the touch-based approach}.  

\subsection{Procedure}
The procedure was adopted from the first evaluation with following changes: After a free testing phase (for every prototype and every function), participants completed a given task set, 
repeated the task with the second prototype. After all tasks were completed the participants filled out the UEQ and a preference questionnaire. For data analysis of performance measures, outliers (distance from median $>$ 1.5 interquartile range) were removed.

\subsection{Tasks}
In total, the participants had to perform 56 tasks: seven for each operation (select, scale, rotate, translate) and prototype (AR and non-AR prototype). The tasks were the same for each prototype to ensure the results are comparable. In every task set for an operation, there was at least one task for each body part of the model, to make sure different conditions such as different object sizes are tested. For the same reason, the tasks were designed to cover different target transformations. For rotation for example, some tasks only required small rotations while others required big or even multiple ones around different axes. 

For every task, the same character model was displayed in the so called T-pose, as can be seen in figure \ref{fig:narScreen}. The T-pose is a typical modeling pose where the character stands on straight legs and the arms point sideways. The size of the camera frustum the model takes up was matched between prototypes. The model consisted of seven body parts: head, neck, two arms, two legs and the torso. For a better distinction of the body parts, they all featured different colors. To ensure that rotations could be seen correctly, they furthermore had a checkerboard texture with a diagonal gradient applied to them. 

At the bottom of the screen a text box displayed the task to accomplish, such as "Select: Head" or "Translate: Green Arm". For body parts that are not unique, such as arms and legs, the color of the associated body part was added to the task text instead of directions such as right and left to prevent confusion and additional cognitive load. For selection tasks, there were no further aids. For other operation tasks however, some further assistance was implemented to allow the user to carry out the task without distractions or dependencies on other interactions. One such assistance was the preselection of the concerned body part. This assistance is important to measure the performance of the operations without influencing it with the performance of the selection interaction. When all task sets had been finished, the application prompted the user to fill out the concluding questionnaire and quit when the user pressed confirm. 

For all tasks except for selection tasks, a semi-transparent duplicate of the affected body part was shown in the same place as the original body part. This duplicate visualizes the transformation objective and will subsequently be referred to as "hint" or "ghost". The ghost is spatially registered in the 3D environment, which can be seen in Figure \ref{fig:narScreen} for the non-AR prototype, and can be occluded by other objects, such as body parts. 

A task was finished by tapping a check mark button (lower left \ref{fig:narScreen}) in Figure. A task could be finished at any time by the user without regard to progress, accuracy or correctness. Whenever a task was finished, information about the performance of the user in this task was stored on the device. This includes the total time needed to fulfill the task, the time needed since the last reset, the count of resets used and the deviation from the target.  For rotation tasks, the angle in degrees between the target and actual rotation is stored as a floating point value. For scale and translation operations, the deviation was calculated for every axis by subtracting the position or scale vector of the ghost object from the according vector of the actual body part object. 

\subsection{Results}
The mean task completion times (TCTs) for the individual tasks were as follows for the AR and non-AR (NAR) prototype. \emph{Selection:} AR: 8.14s (sd=3.41) NAR: 2.67s (sd=0.92). \emph{Translation:} AR: 38.55s (sd=21.65), NAR: 17.34s (sd=8.97). \emph{Rotation:} AR: 75.96s (sd=61.28), NAR: 79.33 (sd=46.49). \emph{Scaling:} AR: 15.64s (sd=6.79), NAR: 15.45s (sd=6.77). For selection and translation, the non-AR prototype was significantly faster as indicated by Wilcoxon signed rank tests (as data was not normal distributed as indicated by Shapiro-Wilk normality tests), but not for rotation and scaling. \emph{Selection}: p$<$0.0001, Z=6.78, Cohen's d=1.38. \emph{Translation}: p$<$0.0001, Z=5.01, d=0.92. \emph{Rotation}: p=0.051, Z=-1.95, d=-0.33. \emph{Scale}: p=0.06, Z=0.52, d=0.088.

\begin{figure}[!htbp]
    \centering
\includegraphics[width=1\columnwidth]{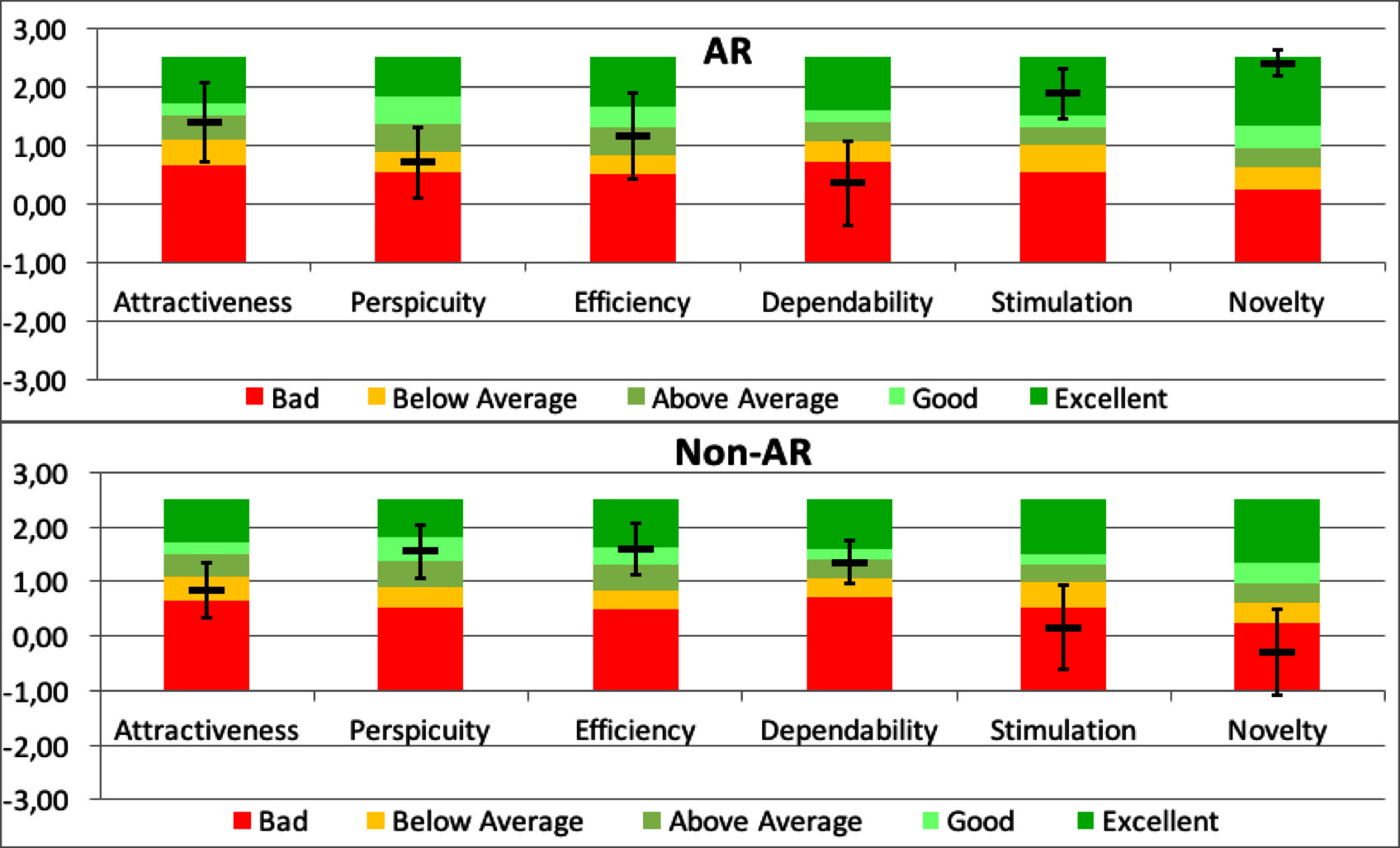}
    \caption{Results of the UEQ for the final TUI-based prototype (top row) and non-AR prototype (bottom row) in the final evaluation.}
    \label{fig:ueqfinal}
    \vspace{-0.5cm}
\end{figure}

The results of the UEQ are shown in Figure \ref{fig:ueqfinal}. 
Wilcoxon signed rank tests indicated statistically significant differences for stimulation (p=0.016, Z=2.40, Cohen's d=1.27, median AR=1.88, median NAR=-0.25) and novelty (p=0.005, Z=2.80, d=1.61, median AR=2.38, median NAR=-0.63), but for no other dimension.


80\% of the partakers explicitly described the TUI-based prototype to be fun to use and five out of ten think it is innovative, exciting and more aesthetically pleasing than the traditional version. 

A problem with the rotation was that 30\% of the users did not fully understand the interaction: instead of drawing circles with the tip of the pen to rotate the object, they only moved it in one dimension, either up and down or left and right, which still worked but probably led to a different user experience than intended. This misunderstanding occurred even though they had time to try out these interactions in the free testing period in which the participants were free to ask for help but were not corrected proactively.   In general, the rotation was said to sometimes require too much effort, especially when standing up for rotations around the Y axis is required, which was even stated to be annoying. 

Some people were surprised by the operations to result in absolute changes instead of relative ones. Five people mentioned that for the rotation interaction and four for the translation. While for some this was only surprising, a few found that annoying, stating by that they are not able to make smaller refinements after an initial manipulation. 
About the MarkerPen itself, four participants stated that the cube to which the markers are applied is too big and gets in the way when being close to the smart phone or table surface. This experience was highly affected by how the user held and interacted with the MarkerPen. While some participants skilfully rotated the pen around or de- and increased the distance to the camera as needed, others interacted with the pen in a less swift and flexible manner, leading to the problems stated above.  The usage of the MarkerPen needed some practice as three participants stated, but after that, one of the participants stated, it is very easy to use. 

 
Three people seemed to have problems with the perception of depth in this environment. Especially in translation tasks, they were repeatedly surprised where they placed the object in depth when viewing it from the side. The same users stated that the translation interaction is annoying, frustrating and too sensitive. Nonetheless, participants highly improved after finding out they could use the shadows of the pen tip and the ghost object as hints to match the depths of the objects. 

To be required to interact also with buttons on the touchscreen of the device was annoying according to three users and they would prefer hardware buttons on the pen itself. Particularly unpleasant was the placement of the buttons to start a new task and the reset button, which were placed in the middle of the screen. This was especially inconvenient for participants who did not use a mounting as they already used both of their hands for holding the device and the MarkerPen.

Camera control was another positively remarked topic. Three people explicitly stated to like the camera controls, mentioning it would be easier and more intuitive than in the non-AR approach. Furthermore, they liked the possibility to manipulate the object and change camera perspective at the same time.

While the prototype as a whole was perceived as intuitive by three people, the rotation interaction did not seem intuitive to the same amount of people, who stated it would require a high cognitive effort. One participant however also remarked that this it felt like "a fun puzzle game". 

The non-AR prototype proved to be intuitive to use. Half of the probands did not even needed instructions. One of those probands however mentioned to have needed the manual for understanding the rotations but did not need it for the other functions. Four of the probands liked being able to work only on one axis without affecting the others as this for example prevented unintended translations in depth. 

80\% stated to prefer the AR prototype. The main reason for that was, that the AR prototype was more exciting and fun to use, as this was stated by six out of eight participants who preferred the AR version. Specifically, they described it as cool, innovative and refreshingly different. Two partakers stated practical reasons for their preference: One user liked being able to manipulate the character and the viewport at the same time. Two others stated the interaction feels more real and direct, creating a better feeling of control. Of the eight users who prefer the AR prototype, five prefer to use it without a smart phone mounting.  Of the two who prefer the non-AR version, one person stated that it is quicker to use. If the editor would be part of a game, the person stated to like to play instead of spending extra time on character editing. The second person explained their preference with the non-AR version being more simple, predictable and stable. 

\subsection{Discussion}

The study indicated that participants found the tangible AR prototype innovative and exciting to use. Although the performance of the prototype was good in terms of hedonic measures and efficiency, further experiments with interaction designs should be made and compared to maximize the usability. The task completion times indicated, while for simple tasks such as selection, the non-ar prototype was significantly faster, their was no significance difference for more complex tasks such as rotation. This might be due to the fact, that in the non-AR prototype rotation can require multiple subsequent actions while in AR (theoretically) a single motion with the MarkerPen is sufficient. Regarding selection tasks, a traditional approach should be considered if the user has a free hand available for interacting with the touch screen, as the selection using tap gestures is significantly more efficient and the users are accustomed to it. A design that allows selection on the touch screen as well as with the marker enriched device could also be a promising way to go, as then people who use a mounting can use their free hand for selection on the touch screen. An interesting general finding was the lower correlation between the satisfaction with the time needed and the actual time passed for the AR prototype. This leads to the assumption, that the actually passed time is a less important factor in the AR version, which might be caused by the participants having more fun than with the non-AR prototype, as people tend to underestimate durations while having fun \cite{thayer1975eye}. Another possible reason might be a different demand in concentration for both prototypes, which could also affect the time perception \cite{jones1976time}. However, the AR version also was perceived as more difficult compared to the non-AR prototype. The reasons for those effects could be investigated in future research to gain further insights into perceived differences between both prototypes. 

Overall, the proposed AR prototype performed  well in the evaluation. While it is indicated that it can't compete with the traditional prototype regarding pragmatic measures such as efficiency and dependability, for which it can however still deliver fairly satisfying results, it performs well in terms of hedonistic measures. This could make it suitable for mobile AR games, for which  efficiency and accuracy might not as important as an innovative and fun interaction design. The presented prototypes only implemented the four basic manipulations selection, rotation, scale and translation. There are further manipulations that might be suitable for AR character editing, such as deformation. 
Furthermore, character editors often implement various other functions, such as changing colours and textures of objects. How well the AR environment and the presented prototype specifically suit such functions and other manipulations should be investigated in future research. While the evaluations tested the performance of individual manipulations in different tasks, it remains unclear how well the prototype would perform in a free setting, where the different manipulations are not isolated from each other and users don’t have to solve tasks. Finally, different types of games will impose different requirements onto a character editor. Further studies should be conducted, to investigate the performance of the presented prototype in different settings, also beyond gaming, e.g., in hedonic oriented touristic \cite{grubert2015utility} or marketing applications. 


\section{Conclusion}
Two AR user interface concepts and according initial prototypes were developed, one based on camera movements, the other on a tangible pen. 
Both proved to be applicable for the task of character editing, however, the MarkerPen-based approach proved to be more entertaining.
This concept was further developed in further design iterations. This prototype was then compared to a traditional non-AR approach. The proposed MarkerPen-based approach performed well in terms of hedonistic measures while also keeping the performance in pragmatic measures such as efficiency on a proficient level. In direct comparison with a traditional approach, most probands preferred the proposed AR prototype for it being more fun and exciting, more direct to control and for more intuitive viewport control.  As AR modeling proved to be more entertaining and exciting than traditional approaches, game developers could use this information to improve the user experience and satisfaction of mobile AR games by adding an AR character editor based on the proposed prototype. Furthermore, this work can serve as base for future research in the fields of AR character editing, AR 3D modeling and tangible user interfaces. 


\bibliographystyle{abbrv}
\bibliography{literature-short}
\end{document}